# Effect of errors on arm asymmetry of doubles

*by*

**Dilip G Banhatti**
[dilip.g.banhatti@gmail.com, banhatti@uni-muenster.de]
School of Physics
Madurai Kamaraj University
Madurai 625021
India

**Abstract / Summary**
Measured values of arm asymmetry parameter $x = (\theta_> - \theta_<)/(\theta_> + \theta_<)$ of a double have appreciable random errors due to errors in positions of radio peaks & of the optically identified galaxy or quasar. These broaden the monotonic decreasing x-distribution $g(x)$. In addition, finite resolution & blending of complex structure leads to errors in recognizing peaks leading to systematic overestimate of x. Thus both random & systematic errors broaden $g(x)$, & consequently broaden the distribution $p(v)$ of derived hotspot separation speed v, & shift its peak to larger v, since $p(v) = -v.g'(v)$.

## Fractional arm difference x

The arm asymmetry of a straight extended extragalactic edge-brightened (i.e., FR2 (Fanaroff & Riley 1974)) double radio source associated with an active galaxy is very usefully specified by the fractional arm difference $x = (\theta_> - \theta_<)/(\theta_> + \theta_<)$ (Banhatti 1979, 1980, 1984/5, Best et al 1995), defined to be the ratio of the difference of the two arms to their sum, so that $0 \leq x \leq 1$, ranging from fully symmetric double (x = 0) to fully asymmetric one (x =1).

## Errors affecting $x = (\theta_> - \theta_<)/(\theta_> + \theta_<)$

The lengths $\theta_>$ & $\theta_<$ of the two arms of a double are determined from the positions of the hotspots & central radio cores or optical identifications, which have random errors of measurement. Moreover, the location of a hotspot is systematically shifted towards inner part of the linear source due to blending with lobe & coarse resolution. Thus, $\theta_>$ & $\theta_<$ are systematic underestimates. Taking true values to be $\theta_> + \Delta$ & $\theta_< + \Delta$ ($\Delta > 0$),
$x_{true} = (\theta_> - \theta_<)/(\theta_> + \theta_< + 2.\Delta) = x/(1 + 2.\Delta/\theta)$ where $\theta_> + \theta_< \equiv \theta$ is the total angular size of the straight double. Thus $x_{true} < x_{measured}$, so that the measured values of x are overestimates. One may take $\Delta$ to be independent of $\theta$ to first approximation. Then $x_{true}$ is most different from $x_{measured}$ for smallest $\theta$ (= overall angular size). For the error on x to be independent of x, x & $\theta$ should be uncorrelated. This indeed seems to be the case, as tested by Teerikorpi's (1986) Table 2.1 & Fig.2, using x vs $\theta$ scatter diagram.

The random errors on x are affected by many factors, & hence may be taken to be Gaussian, equally likely to be positive or negative.

Banhatti (1979, 1980, 1984/5) uses the distribution g(x) of x to derive the distribution p(v) of hotspot separation speeds v in an intrinsically symmetric model, where the asymmetry is fully ascribed to double source orientation out of sky plane & (consequent) light-travel time differences for radiation emanating from the two hotspots. He finds that **p(v) = - v.g'(v)** where prime denotes differentiation. From observed samples of doubles, the function g(x) is found to be decreasing from a maximum at x = 0 to zero at x = 1. Considering the x-distribution g(x) as a histogram, the number of sources going out of a bin due to random errors in x is proportional to the bin population. Since the bins decrease in population toward large x, more x-values will shift toward larger x at every x, so that g(x) widens due to random errors in x.

Thus both systematic & random errors in x cause widening of the x-distribution g(x). This leads to a corresponding widening of the v-distribution p(v), & a shift of its peak to a higher value. Thus the true p(v) is narrower & is peaked at smaller v than measured (Banhatti 1984/5).

[*Readers are welcome to email author with suggestions / comments / queries.*]